\documentclass{article}
\usepackage{graphicx,comment} 
\usepackage[margin=1in]{geometry}
\usepackage{float}
\usepackage{makecell}
\usepackage[mathscr]{euscript}

\usepackage[utf8]{inputenc}
\usepackage{comment}
\usepackage{amsmath}
\usepackage[thinc]{esdiff}
\usepackage{mathtools}
\usepackage[T1]{fontenc}
\usepackage{lmodern} 
\usepackage{blindtext}

\usepackage{diagbox}
\usepackage{pdflscape}
\usepackage{booktabs}
\usepackage{caption}
\usepackage{caption}

\usepackage{multirow,tabularx}
\newcolumntype{Y}{>{\centering\arraybackslash}X}
\renewcommand\arraystretch{2}

\title{Sensitivity analysis for matching on high-dimensional predictors: A case study of racial disparity in US mortality}
\author{Marina Hernandez, Ciprian Crainiceanu}
\date{Johns Hopkins University Bloomberg School of Public Health, Baltimore, MD}

\RequirePackage[table]{xcolor}
\RequirePackage{booktabs}
\RequirePackage{multirow}

\RequirePackage{caption}
\begin{document}
\captionsetup{textfont=normalfont}
\maketitle

\begin{abstract}
Matching on a low dimensional vector of scalar covariates consists of constructing groups of individuals in which each individual in a group is within a pre-specified distance from an individual in another group. However, matching in high dimensional spaces is more challenging because the distance can be sensitive to implementation details, caliper width, and measurement error of observations. To partially address these problems, we propose to use extensive sensitivity analyses and identify the main sources of variation and bias. We illustrate these concepts by examining the racial disparity in all-cause mortality in the US using the National Health and Nutrition Examination Survey (NHANES 2003-2006). In particular, we match African Americans to Caucasian Americans on age, gender, BMI and objectively measured physical activity (PA). PA is measured every minute using accelerometers for up to seven days and then transformed into an empirical distribution of all of the minute-level observations. The Wasserstein metric is used as the measure of distance between these participant-specific distributions. 

\end{abstract}

\section{Introduction}
Matching on multilevel, high-dimensional predictors is a largely unaddressed problem in statistics. Identifying a measure of distance between high-dimensional measurements, determining what distance threshold to use, and examining the within- and between- person variability of these distances are important, open problems. In this paper, we explore the potential pitfalls of high-dimensional matching and propose extensive sensitivity analyses as a partial solution. The goal of these analyses is to identify the parameters and choices that have the highest impact on the resulting analyses. To illustrate these points, we focus on estimating the hazard ratio of all-cause mortality in the US among African Americans compared to Caucasian Americans \cite{benjamins_comparison_2021,beydoun_racial_2016,borrell_racialethnic_2010,luo_mortality_2021}. Data are obtained from the National Health and Nutrition Examination Survey (NHANES 2003-2006) and African American individuals were matched to Caucasian Americans based on on age, gender, BMI, and objectively measured physical activity (PA). For the purpose of this paper, PA is a time series of minute-level summaries of PA intensities (expressed in Activity Counts \cite{leroux_organizing_2019}) for up to seven days per person. These time series are transformed into distributions \cite{ghosal_predicting_2023,ghosal_distributional_2023,ghosal_scalar_2022,matabuena_distributional_2023} and the Wasserstein distance is used to quantify the distance between the resulting distributions. We illustrate that decisions about the computation of the distance as well as the caliper width have a substantial impact on identifying the specific subgroups and, ultimately, on the estimated hazard ratio and its associated variability.  This problem is likely not specific to the Wasserstein distance.  

We now provide a short introduction to matching and the Wassertein distance for distributions and discuss practical implications in NHANES. The rest of the paper then proceeds with the proposed sensitivity analysis and an in-depth discussion of the effects of day-to-day variation (biological measurement error) in PA on estimators of all-cause hazard ratio of mortality. The paper ends with a short discussion with recommendations.

\section{Background}

\subsection{Matching}\label{sec:matching}

Historically, matching has been used for observational studies in low-dimensional settings, where individuals are matched on a few covariates, such as age, sex, and BMI \cite{stuart_matching_2010}. Its stated purpose is to improve the balance of the covariates used for matching, though it remains unclear what happens to the other observed and unobserved covariates. For example, \cite{king_why_2019} showed that matching on observed covariates can lead to a worse balance of the unobserved covariates, which could lead to bias. This is an under-recognized problem in statistics and very few methods have been proposed to identify such biases.

While sensitivity analyses have been discussed in the context of matching \cite{ rosenbaum_sensitivity_1991,rosenbaum_sensitivity_2014,rudolph_using_2018,stuart_matching_2010}, there is currently no guidance on how to conduct sensitivity analyses, which parameters to investigate, and what inferential results to report. {\it Undoubtedly, sensitivity analyses raises the costs of data analyses, but it is the lowest possible analytic standard for matching analyses.}

\subsection{Wasserstein Distance}\label{sec:distance}

The focus of our paper is on matching with high dimensional data. The problem was motivated by large observational studies that collected objectively measured physical activity (PA) data every minute for up to seven days (up to $7\times 1{,}440=10{,}080$ observations per person). Summaries of these data were shown to be as predictive as or more predictive than age for all-cause and cardiovascular mortality \cite{ledbetter2022,leroux_quantifying_2020,smirnova_predictive_2020}. Thus, when conducting matching, it makes sense to consider matching on PA in addition to traditional covariates such as age, gender, and BMI. As PA is high dimensional, we need to identify a distance between PA of all study participants, while taking into account that some individuals have fewer days of data (as few as three and as many as seven).

For the purpose of this paper we decided not to match directly on the time series of minute-level activity counts. Instead, for each participant, we transform the observed time series of minute-level AC observations into an empirical distribution \cite{ghosal_distributional_2023,ghosal_scalar_2022,matabuena_distributional_2023,ghosal_predicting_2023}. The distribution of the observed time series has the advantages that: (1) it can be estimated with a different number of observations per person; and (2) contains information about time in every range, the mean, quantiles, the standard deviation, and many other summaries of PA that one might consider for matching. It has the disadvantage that it loses information regarding the timing of PA, including when it occurred during the day and on which day.

The Wasserstein distance can then be used for any pair of estimated PA distributions. We define the theoretical Wasserstein distance as follows: if $Q_1(p)$ and $Q_2(p)$ for $p\in[0,1]$ are the theoretical quantile functions of two distributions, the Wasserstein distance is the L2 distance between these quantiles, 
\begin{equation}D_W(Q_1,Q_2)=\sqrt{\int_0^1 \{Q_1(p)-Q_2(p) \}^2dp}\;.
\label{eq:Was_integral1}
\end{equation}
However, this is where the theoretical rubber meets the practical road. Indeed, neither $Q_1(\cdot)$ nor $Q_2(\cdot)$ are available at every $p\in[0,1]$. They can only be estimated based on a finite sample, at a finite number of values, $p_j$, with extreme estimated quantiles (when $p$ is close to zero or one) being subject to more error. Thus, we instead use the following approximation
\begin{equation}
\widehat{D}_W(\widehat{Q}_1,\widehat{Q}_2)\approx\sqrt{\frac{1}{J}\sum_{j=1}^J\{\widehat{Q}_1(p_j)-\widehat{Q}_2(p_j)\}^2}\;,
\label{eq:Was_Riemann2}
\end{equation}
which is the Riemann sum approximation to the integral in \eqref{eq:Was_integral1}. The hats indicate that quantities are estimated from the data and $p_j$ is an equal grid of points of length $J$ in $[0,1]$. Empirical quantiles are calculated in probability increments of $1/J$. 

\subsection{Motivating Example: NHANES Physical Activity Data}\label{sec:example}

To begin illustrating the challenges of matching on high-dimensional PA, we first need to understand the context and structure of the NHANES 2003-2006 data. Objectively measured PA data were collected in the free-living environment on participants who were asked to wear a hip-worn PA monitor for seven consecutive days. The resulting data were summarized at the minute-level in a proprietary measure called activity count (AC); for more details on the NHANES objectively measured PA data, see \cite{leroux_organizing_2019}. Thus, for each participant, PA measurements were obtained at every minute of the day for up to seven consecutive days ($1{,}440$ minutes of AC per day), and days were discarded for individuals if NHANES assessed that the data on that day were not of ``good quality''. We define a ``good day'' as one with the following properties: (1) it has an estimated wear time of over 10 hours, (2) the data are calibrated, and (3) the data are deemed reliable by NHANES. 

\begin{figure}[H]
    \centering
    \includegraphics[width=1\textwidth]{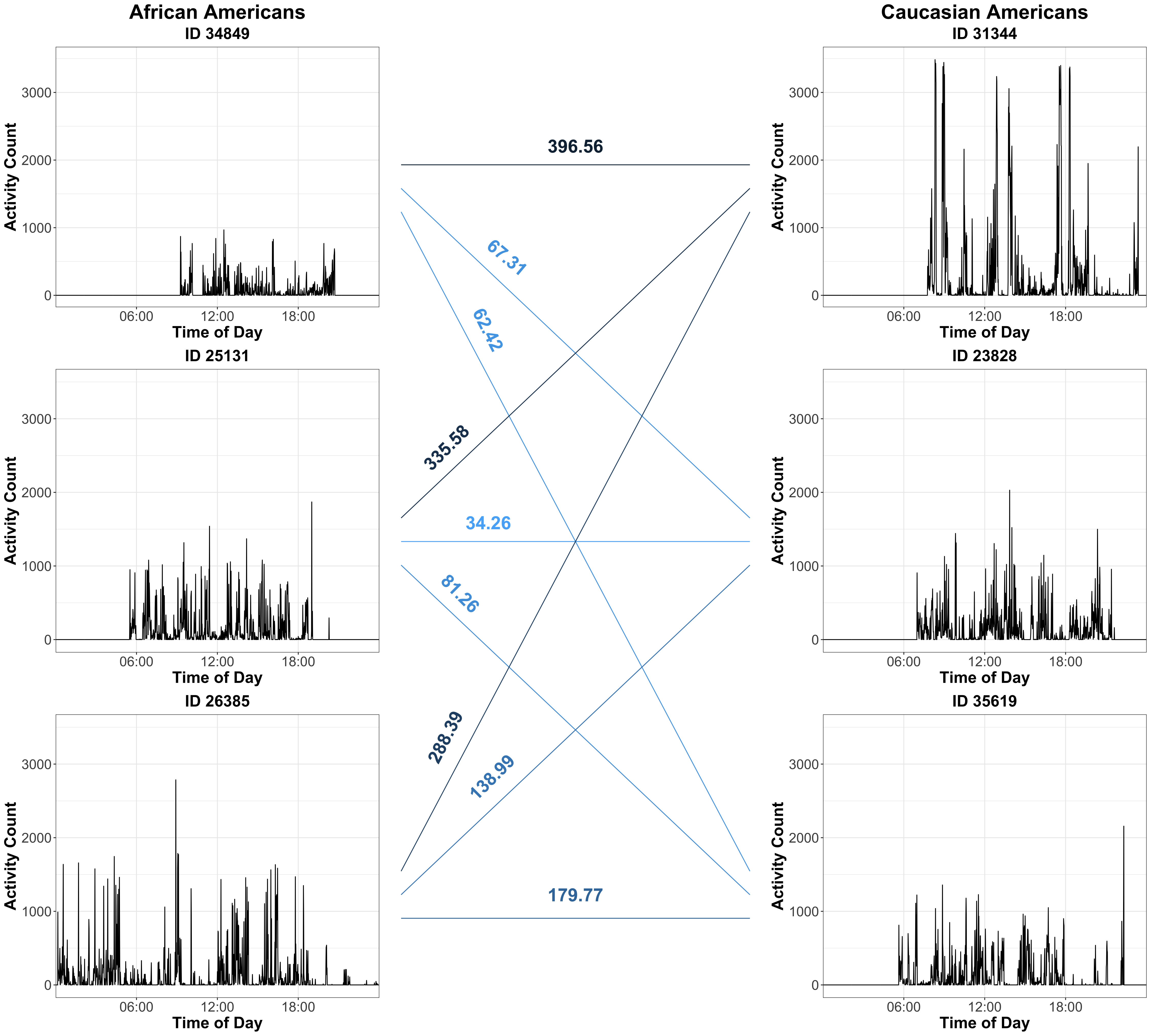}
    \caption{A single day of physical activity for three African Americans (left panels) and three Caucasian Americans (right panels), and the Wasserstein distances, computed between individuals using all their PA data across days, displayed on the lines}
    \label{fig:mesh1}
\end{figure}

Figure~\ref{fig:mesh1} presents PA data for six individuals, including three African Americans (left panels) and three Caucasian Americans (right panels). For illustration purposes, each panel displays a single, complete day of the corresponding participant's PA pattern, though each individual has between $3$ and $7$ ``good days''. The y-axis is kept the same for all study participants to ensure that observations are comparable across participants. From these data we can see that: (1) PA measurements have substantial within-day and between-person variability; (2) the timing of PA is not synchronized across individuals and their peak activity as well as sleep periods occur at different times; (3) data exhibit substantial skewness; (4) the unit of measurement, Activity Count, is not easy to understand or translate into actionable information; and (5) the complexity and size of the data makes traditional analyses and decisions difficult.

The Wasserstein distance is displayed in Figure~\ref{fig:mesh1} between every pair of study participants as numbers on the lines. Recall that this distance is computed as described in \ref{sec:distance}. Smaller distances suggest that the two individuals have a more similar pattern of PA, at least in terms of overall time spent at different levels of PA. For example, the distance between the study participant shown in the top left panel (African American, ID 34849) and the study participant in the top right panel (Caucasian American, ID 31344) is $396.56$, while the distance to the study participant in the middle right panel (Caucasian American, ID 23828) is $67.31$. The smaller PA distance to the second Caucasian American seems to be consistent with the visual inspection of the three panels (top left, top right and middle panels). 

This inspection of the Wasserstein distances in Figure~\ref{fig:mesh1} illustrates some potential problems with matching via distances. First, what is the exact meaning of the resulting distance between the PA patterns of any two individuals? Second, how should one decide whether a distance is large or small and when can we say that two individuals are close enough to be matched on PA patterns? Third, should matching be conducted without accounting for the number of study participants who are matched? Fourth, how should the additional days of activity data for each individual be handled? Fifth, how should the variability of point estimators given these choices be reported? Given all these difficulties, the problem of matching on PA patterns looks increasingly daunting. 

To begin to understand where the distances displayed in Figure~\ref{fig:mesh1} come from and what they may mean in terms of the original data, we now explore how these distances were derived. Table~\ref{tab:quants} presents some of the quantiles of physical activity on the activity counts scale for three individuals from Figure~\ref{fig:mesh1} (IDs 34849, 31344, and 23828). Here, for example, $Q.80$ denotes the estimated $0.80$ probability quantile. For study participant with ID 34849, $Q.80=34.00$ can be interpreted as $80$\% of this participant's minute-level activity counts were below $34.00$ and $20$\% were above.

\begin{table}[htbp]
    \renewcommand{\arraystretch}{1.3} 
    \setlength\belowcaptionskip{8pt}    
    \centering
    \caption{}
    \begin{tabular}{lrrrrrrr}
        \toprule

ID & Q.50 & 
 Q.60& Q.70 & Q.80 & Q.90 & Q.95 & Q.99 \\
        \midrule

 \rowcolor[rgb]{ .906,  .902,  .902} \multirow{1}[0]{*}{\textsc{34849}} &0.00  &  0.00  & 2.00 & 34.00 & 192.00 & 441.00 & 940.43\\ [7 pt]

       \rowcolor[rgb]{ 1,  1,  1}\multirow{1}[0]{*}{\textsc{31344}} & 0.00 &   2.00   &24.30 & 138.00  & 627.00 & 1174.00 &3259.68 \\ [7 pt]

 \rowcolor[rgb]{ .906,  .902,  .902}  \multirow{1}[0]{*}{\textsc{23828 }} & 0.00   & 12.00  & 57.00 & 150.00 & 316.00 & 500.00 & 1091.84 \\ [7 pt]

        \bottomrule
        
\end{tabular}%
\label{tab:quants}%
\end{table}%

Let us now explore the quantile distances between the pairs (34849, 31344) and (34849, 23828). Rather than examining the entire Riemann sums, we focus only on the following portion
$$\sum_{j}\{\widehat{Q}_1(p_j)-\widehat{Q}_2(p_j)\}^2\;,$$
where $p_j\in\{0.5,0.6,0.7,0.8,0.9,0.95,0.99\}$. For the pair (34849, 31344), we obtain
\[(0.00-0.00)^2+(0-2)^2+...+(441-1174)^2+(940.43-3259.68)^2=6{,}116{,}752\;.\]
For the pair (34849, 23828), we obtain
\[(0.00-0.00)^2+(0-12)^2+...+(441-500)^2+(940.43-1091.84)^2=58{,}407\;.\]
A quick inspection of these two values reveals something unexpected; both sums are strongly dominated by the differences in the highest quantiles. Indeed, the values in $Q.95$ and $Q.99$ are much larger than in any of the other quantiles. 

In theory, one would be interested in making $J$ as large as possible to improve the approximation of the integral~(\ref{eq:Was_integral1}) by the Riemann sum ~(\ref{eq:Was_Riemann2}). This notion is misguided because, when spacing between probabilities is reduced, the effects of high quantiles is increased. Recall that the total number of PA observations for one person is, at most, $1{,}440$ (minutes)$\times 7$ (days)$  =10{,}080$ observations. Therefore, when $J$ increases above $1{,}000$, the number of observations in each bin decreases rapidly. The effect is that quantiles are barely changing for a small increase in probability. Even though this may appear to be what we want, we have just shown that the sums are dominated by the high quantiles. This could be especially problematic because the tails of the distributions are much more sensitive to noise. 

\section{Sensitivity Analysis To High-Dimensional Matching}\label{sec:sensitivity}

\subsection{Matching Setup}
To demonstrate the challenges and potential pitfalls of high-dimensional matching described in Section \ref{sec:example}, 
we use data from the NHANES waves 2003-2004 and 2005-2006 to examine the hazard ratios for all-cause mortality of African Americans relative to Caucasian Americans after matching. A combined total of $2{,}486$ participants in these waves met the criteria for inclusion in our analyses; for a detailed inclusion/exclusion chart, see Figure \ref{fig:flowchart} in the supplementary materials. In particular, $1{,}879$ and $607$ participants identified themselves as ``White'' or ``Black'', respectively, were over the age of $50$, had good quality accelerometry data, and were not missing the covariates or outcome of interest.

Then, to obtain estimates for the hazard ratio, we implement the following functional linear Cox regression model \cite{cox_regression_1972,cui_functional_2024,gellar_cox_2015,kong_flcrm_2018}, where we quantify the hazard ratio for the race variable by the parameter $\gamma_1$:  

\begin{equation}
    \log\{h_i(t)\}=\log\{h_0(t)\}+\gamma_1*{\rm Race}_i+\gamma_2*{\rm Age}_i+\gamma_3*{\rm Gender}_i+\gamma_4*{\rm BMI}_i+\int_\mathscr{S} X_i(s)\beta(s)ds\;.
    \label{eq1}
\end{equation}
Here, $h_0(t)$ is an unspecified baseline hazard function, $X_i(s)$ is the average physical activity for study participant $i$ at time of the day $s$ over all available days, and $\beta(s)$ is a functional coefficient.

For reference, if we were to apply model~\eqref{eq1} to the total $2{,}486$ participants without matching, the estimated hazard ratio is 
$\exp{(\widehat{\gamma}_1)}=1.104$ with a $95$\% confidence interval of ($0.92,1.36$). Although the result is not statistically significant, the point estimator suggests a $10$\% increase in the hazard of mortality for African Americans relative to Caucasian Americans, which is consistent with multiple literature reports \cite{benjamins_comparison_2021,beydoun_racial_2016,borrell_racialethnic_2010,luo_mortality_2021} based on much larger data sets. 

We compare this estimate based on the total sample with results obtained from one-to-one matching. To be specific, we match African Americans to Caucasian Americans on age ($\pm$ 3 years), BMI ($\pm$ $2$ kg/m$^2$), gender, and PA \cite{leroux_quantifying_2020,smirnova_predictive_2020}. The variable name ``gender'' used in the data set is taken directly from the framing of the questions in NHANES, and is not intended to conflate sex and gender.  As mentioned in Section \ref{sec:distance}, we match on PA as follows: for each individual, a single PA distribution is created using all ``good days'', and the Wasserstein distance is calculated between the quantile functions for each pair. If multiple Caucasian Americans can be matched on these variables to an African American, then we randomly sample one of these individuals without replacement (that is, if one Caucasian American was chosen to be in the matched group, they cannot be sampled again.) 

This matching mechanism depends on multiple choices, including the calipers for age, gender, BMI, and PA, and the parameters used for the calculation of the Wasserstein distance. Therefore, each such set of parameters will create a different subset of the original data, which may lead to different or even contradictory results. To address this problem and those discussed in Section \ref{sec:matching}, we propose to conduct extensive sensitivity analyses on the following set of tuning parameters:

\begin{enumerate}
    \item $C$, the caliper of the Wasserstein distance;
    \item $J$, the number of probability increments for computing distance;
    \item $I$, the interval on which the distance is computed.
\end{enumerate}

For each matched population generated by a set of these parameters, we fit model~\eqref{eq1}. Recall that, even when the matching parameters are fixed, the procedure for choosing the matched data will provide different subsets. The reason is that each African American participant may have more than one possible Caucasian American match. Since we are conducting one-to-one random matching without replacement, these matches will differ from one data set to another. To account for this variability, the matching process is repeated $30$ times which yields $30$ different estimators of $\exp(\gamma_1)$.

\subsection{Results}

\newcolumntype{L}[1]{>{\raggedright\arraybackslash}p{#1}}
\newcolumntype{C}[1]{>{\centering\arraybackslash}p{#1}}
\newcolumntype{R}[1]{>{\raggedleft\arraybackslash}p{#1}}

\begin{landscape}

\begin{tabular}{L{1cm}|C{1cm}C{1cm}C{1cm}C{1cm}C{1cm}C{1cm}C{1cm}C{1cm}C{1cm}C{1cm}C{1cm}C{1cm}C{1cm}C{1cm}C{1cm}C{1cm}}
\toprule
$J$ & 1 & 3 & 4 & 9 & 19 & \multicolumn{2}{c}{99}
&
\multicolumn{3}{c}{199} &
\multicolumn{3}{c}{999} &
\multicolumn{3}{c}{1999}  \\
\cmidrule(r){0-5}\cmidrule(r){7-8}\cmidrule(l){9-11} \cmidrule(l){12-14} \cmidrule(l){15-17}  
  \diagbox[width=1.1 \textwidth/13 \relax, height=1cm]{ $C$ }{$I$}& * & * & * & * & * &** & [.05,.95]   & [0,1]&[.01,.99]  &[.05,.95]   & [0,1]&[.01,.99]  &[.05,.95] & [0,1]&[.01,.99]  &[.05,.95] \\
\toprule
 \hline

  10 & 1.14 (576) & 1.16 (484) & 1.13 (412) & 1.02 (224) & 1.02 (96) & 1.62 (20) & 1.42 (66) & 1.78 (11) & 2.43 (25) & 0.88 (139) & NA  & 1.82 (28) & 0.93 (145) & NA  & 1.84 (28) & 0.93 (145) \\ 
  20 & 1.16 (579) & 1.14 (533) & 1.12 (497) & 1.09 (403) & 1.21 (291) & 1.05 (141) & 1.09 (242) & 1.33 (106) & 1.13 (149) & 1.2 (336) & 1.68 (66) & 1.12 (160) & 1.19 (343) & 1.44 (56) & 1.14 (161) & 1.17 (345) \\ 
  30 & 1.13 (581) & 1.12 (552) & 1.16 (530) & 1.11 (479) & 1.17 (417) & 1.2 (264) & 1.16 (362) & 1.23 (228) & 1.26 (275) & 1.09 (446) & 1.26 (172) & 1.28 (289) & 1.1 (449) & 1.23 (159) & 1.22 (289) & 1.08 (450) \\ 
  40 & 1.14 (583) & 1.12 (560) & 1.14 (550) & 1.13 (519) & 1.19 (472) & 1.32 (355) & 1.07 (435) & 1.29 (320) & 1.31 (367) & 1.15 (498) & 1.3 (270) & 1.28 (374) & 1.14 (497) & 1.29 (258) & 1.31 (375) & 1.14 (497) \\ 
  50 & 1.13 (586) & 1.15 (567) & 1.19 (563) & 1.12 (538) & 1.14 (514) & 1.27 (422) & 1.15 (479) & 1.27 (396) & 1.19 (432) & 1.15 (531) & 1.27 (351) & 1.14 (443) & 1.17 (530) & 1.26 (345) & 1.16 (442) & 1.16 (530) \\ 
  100 & 1.14 (588) & 1.17 (584) & 1.16 (579) & 1.16 (569) & 1.18 (561) & 1.18 (540) & 1.17 (555) & 1.17 (532) & 1.17 (541) & 1.18 (565) & 1.13 (524) & 1.17 (542) & 1.16 (565) & 1.14 (521) & 1.14 (542) & 1.19 (566) \\ 
  200 & 1.17 (590) & 1.15 (589) & 1.16 (588) & 1.16 (585) & 1.2 (580) & 1.17 (575) & 1.21 (577) & 1.22 (574) & 1.2 (576) & 1.19 (583) & 1.2 (572) & 1.23 (576) & 1.2 (583) & 1.17 (572) & 1.23 (576) & 1.19 (583) \\ 
  300 & 1.12 (590) & 1.16 (590) & 1.14 (589) & 1.15 (588) & 1.17 (586) & 1.18 (584) & 1.16 (584) & 1.2 (581) & 1.15 (583) & 1.16 (587) & 1.2 (581) & 1.2 (583) & 1.13 (587) & 1.16 (581) & 1.14 (584) & 1.14 (587) \\ 
  400 & 1.14 (589) & 1.14 (589) & 1.13 (589) & 1.12 (589) & 1.16 (588) & 1.17 (586) & 1.15 (586) & 1.18 (586) & 1.13 (586) & 1.11 (588) & 1.14 (586) & 1.13 (587) & 1.17 (588) & 1.15 (585) & 1.14 (587) & 1.14 (588) \\ 
  \hline
\bottomrule
\end{tabular}
\captionof{table}{Average estimated hazard ratios and average number of matched pairs in parentheses across 30 independent repetitions of the matching process for each combination of C, the caliper of the Wasserstein distance, J, the number of probability increments for computing distance, and I, the interval on which the distance is computed}\label{maintable}
\captionof*{table}{* All intervals produce same results \hspace{1cm} ** [0,1] and [0.01,0.99] produce same results}
\end{landscape}

Table~\ref{maintable} displays the average estimated hazard ratios, $\exp(\hat{\gamma}_1)$, computed from $30$ independent iterations of the matching process for various combinations of the tuning parameters: $C$, $J$, and $I$. Below the average hazard ratio, we also provide the average number of matched pairs in parentheses. For example, for $C=20$ (the caliper distance between densities) and $J=99$ (the number of terms used in the Riemann sum approximation) and $I=[0.05,0.95]$ (the interval for integration of the Wasserstein distance), the average estimated hazard ratio is $1.09$ with an average of $242$ matched pairs. 

Overall, the estimated hazard ratios tend to exceed $1$, indicating a heightened risk of mortality among African Americans compared to Caucasian Americans. However, these hazard ratios exhibit substantial differences in their value and provide contradictory results. For instance, when using the tuning parameters $C=10$, $J=199$, and $I=[0.01,0.99]$, the estimated hazard ratio is $2.43$, based on $25$ matched pairs. In contrast, using the same $C$ and $J$ but considering the interval $[0.05,0.95]$, the hazard ratio drops to $0.88$, based on $139$ matched pairs. This demonstrates that even a small change in the interval used for the Riemann sum approximation can lead to substantial changes in the size of the matched population and, more importantly, in the interpretation of the estimator of the hazard ratio. The change in the estimated hazard ratio may indicate that matching on observed covariates may induce bias in point estimators due to unmeasured (or not included) confounders. Another possible explanation for this difference could be that the Wasserstein distance is highly sensitive to extreme quantiles, which can be heavily affected by noise and sample size. In Section \ref{sec:example}, we have shown that this may be the case; in this section, we show that this has practical consequences. 

Although there may be considerable variability among the hazard ratios, Table~\ref{maintable} demonstrates that as the caliper $C$ surpasses $200$, a stabilization occurs irrespective of the number of terms utilized, converging to about $1.15$ with approximately $590$ matched pairs. This implies that around $97\%$ of the $607$ African American individuals have been successfully matched, while the remaining $17$ African Americans could not be simultaneously matched based on age, gender, and BMI. Consequently, for a large enough caliper $C$ for physical activity (as indicated in the last rows of Table~\ref{maintable}), the matching process is less affected by physical activity and is instead driven by age, gender, and BMI. In fact, the resulting hazard ratio for matching on just age, gender, and BMI and implementing model \ref{eq1} is $1.12$, which is close to the hazard ratios reported in the last rows of Table~\ref{maintable}. Thus, the stabilization of point estimators for larger calipers is a manifestation of including almost all African Americans in the analysis and should not be interpreted as a measure of how appropriate the model estimators are. Since almost all African Americans are matched at a caliper of $200$, it becomes evident that this caliper is too large for effective matching, which gives us a sense of what is close and far in terms of the Wasserstein distance; distances exceeding this value would imply extremely different physical activity profiles. This finding underscores the necessity to examine the effects of a range of values for the tuning parameters on the results, not only individually, but jointly. 

Recall that the columns in Table~\ref{maintable} represent the number of terms in the Riemann sum approximation to the theoretical Wasserstein distance and the subcolumns represent the interval of integration. For example, $J=9$ corresponds to the distance $\widehat{D}_W(\widehat{Q}_1,\widehat{Q}_2)$, using the deciles of the distribution. The interval parameter here is represented by an asterisk because using deciles does not affect the interval on which we are conducting the approximation; indeed, the deciles correspond to the same values of probabilities in all intervals: $[0,1]$, $[0.05,0.95]$, and $[0.01,0.99]$. Although there is no clear convergence in the overall hazard ratios or number of matches as $J$ increases and for smaller $C$, a pattern emerges in the first six columns denoted by the asterisks. The number of matches decreases substantially and consistently with increasing $J$. For example, when $C=20$, there are more matches when using the median, $J=1$ ($579$), or quartiles, $J=3$ ($533$), than when using the deciles, $J=9$ ($403$), with implications on the estimated hazard ratio as well. This extreme sensitivity to the choice of the number of quantiles is rather unexpected and illustrates the importance of sensitivity analyses. 

Let us now investigate what happens when the number of quantiles used in the Riemann sum increases beyond $99$. For simplicity, we focus on the interval $[0,1]$ and compare results for $J$ equal to $99$ (subcolumn $6$), $199$ (subcolumn $8$), and $999$ (subcolumn $11$). As $J$ increases, the number of matches decreases substantially, and the hazard ratios vary as well. For example, for $C=20$, there are $141$ matched pairs if we use $99$ quantiles (centiles), but there are only $66$ matched pairs if we use $999$ quantiles. Furthermore, notice that for the interval $[0,1]$ and caliper $10$, two NA's appear as $J$ increases because there were not enough matches to run the model. This behavior can be observed throughout the table, though it becomes less obvious for large calipers, $C$, when almost all African Americans are matched. This is likely due to the fact that, as $J$ increases, the Wasserstein distance becomes more dependent on the extreme quantiles. This could be a problem as the extreme quantile estimators are more likely to be affected by error especially in finite, though large, samples. Although these findings fall short of constituting proof, they suggest the importance of conducting sensitivity analyses when employing high-dimensional matching.

Next, we investigate the differences between the intervals $[0.01,0.99]$ and $[0.05,0.95]$ used for the Riemann sum approximation. We have seen that the number of matches decreases as the number of quantiles in the Riemann sum, $J$, increases using the interval $[0,1]$. However, this trend does not hold as $J$ exceeds $99$ for the intervals of $[0.01,0.99]$ and $[0.05,0.95]$. Instead, the number of matches for both the interval $[0.01,0.99]$ and $[0.05,0.95]$ begin to increase as $J$ increases beyond $99$. For instance, for a caliper of $20$ and the interval $[0.01,0.99]$, there are $141$ matched pairs at $J=99$, and $161$ matched pairs at $J=1999$. A plausible explanation for this could be that as $J$ increases, more quantiles from the central part of the physical activity distribution are utilized in the distance equation, which may reduce distances between participants. This is plausible, as the $[0.05,0.95]$ interval tends to correspond to more matches than the $[0.01,0.99]$ at a fixed caliper, $C$. We conclude that, for a fixed $C$, and a fixed $J$, the interval $[0.05,0.95]$ yields the highest number of matches, followed by $[0.01,0.99]$, and lastly the $[0,1]$ interval. Larger calipers are needed for the $[0,1]$ interval to produce the same number of matches as the other intervals at a fixed $J$. 

In addition to the number of matches in the intervals $[0.01,0.99]$ and $[0.05,0.95]$ as $J$ increases beyond $99$, we now examine the hazard ratios. To illustrate, consider the caliper $C=30$, and $J=199$, $999$, and $1{,}999$. The corresponding hazard ratios for the interval $[0.01,0.99]$ are $1.26,1.28,1.22$, respectively, which are consistent and hover around $1.25$. In contrast, the hazard ratios for the interval $[0.05,0.95]$ are $1.09,1.10,1.08$, respectively, which are approximately $1.10$. Thus, the hazard ratios are quite similar within an interval type as $J$ increases beyond $99$. The same stabilization occurs for other choices of caliper, though the point estimators for hazard ratios continue to vary substantially across intervals. Indeed, even minor changes of the integration limits (e.g., from $[0.05,0.95]$ to $[0.01.0.99]$) leads to substantial changes in the estimates. It is difficult to assess how these differences would be influenced by the size and characteristics of other data sets. 

In summary, the hazard ratios and number of matched pairs are highly sensitive to the choice of the caliper, $C$, integration interval, $I$, and number of terms in the Riemann sum, $J$. In practice, these tuning parameters are chosen without conducting a sensitivity analysis, which substantially reduces the scope of the analysis. Our results indicate that this could be misleading and even abused in practice by simply running analyses with slightly different tuning parameters. While a sensitivity analysis does not solve the problem of finding the ``best estimator'', it provides much needed transparency to the matching process. Indeed, our sensitivity analysis revealed that the process of matching is akin to choosing a subset of the data on which to run the analysis. Recognizing this process as subgroup analysis without knowing the subgroups a-priori is important for how matching is conducted and understood. These problems are particularly relevant in the context of high-dimensional matching, where the idea of distance, its exact meaning, and the definition of closeness are far from being elucidated. 

\subsection{Measurement Error}\label{subsec:meas_error}
The results in the previous section have shown that analyses can be highly sensitive to the choice of caliper, $C$. Unfortunately, determining a value for an appropriate caliper is challenging. In this section, we attempt to inform the selection of $C$ by taking advantage of the fact that we have multiple days of data for each individual. This will address two difficulties that are often ignored in high-dimensional distance-based analyses: (1) recognizing that distances are often subject to substantial measurement error induced by, for example, measurement or biological variability; and (2) introducing the within-subject distances as a measure of distance heterogeneity as well as a reference for what is close or far. Certainly, if the within-person distances hover around $10$, it could be reasonable to consider a match to be someone whose distance to this person is approximately $10$. While this is not a perfect solution, it does provide a step forward in terms of defining closeness in an otherwise very complex space.

\begin{figure}[H]
    \centering
    \includegraphics[width=1\textwidth]{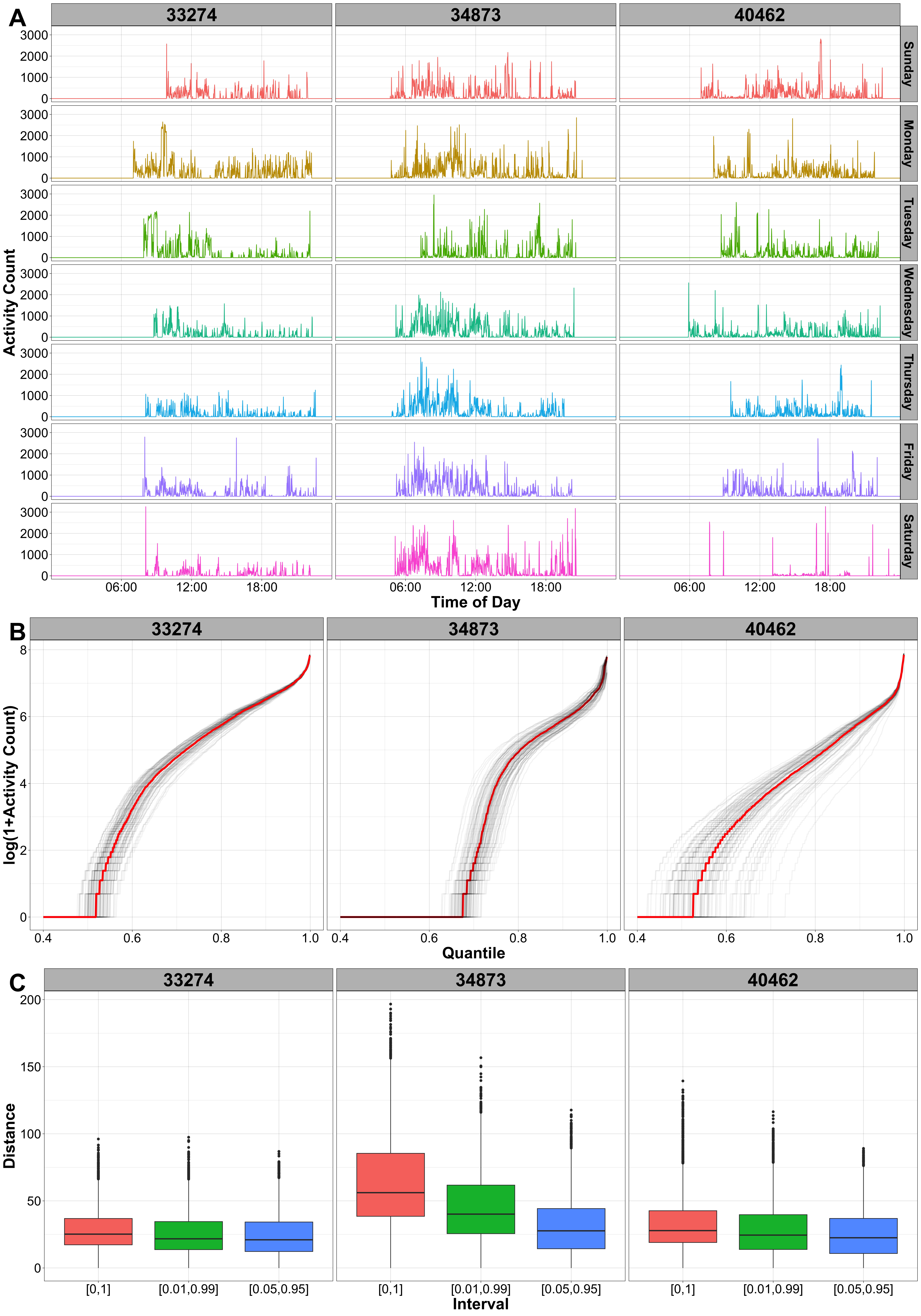}
    \caption{Panel A) Midnight to midnight physical activity for three individuals across seven days. Panel B) 100 quantile functions constructed from resampled PA days and the true estimated quantile function in red. Panel C) Within-person Wasserstein distances between all 100 curves for three interval types.}
    \label{fig:meas}
\end{figure}

In Figure~\ref{fig:meas}, the impact of within-person variability, or measurement error, on high-dimensional physical activity matching is depicted for three individuals. Each sub-panel of Figure~\ref{fig:meas}-(A) represents the data observed at the individual level across seven days. To obtain an estimator of the distribution, these data are stitched together within each individual and across their 7 days, and the empirical quantiles are obtained from this joint vector. Because we have multiple days of observations, the process can be repeated by conducting a bootstrap of days and recalculating the distribution for each bootstrap sample. This process mimics the natural day-to-day biological variability observed in objectively measured physical activity. The panels in Figure~\ref{fig:meas}-(B) display the estimated quantile functions after conducting $100$ bootstrap re-samples within each participant. The spread of gray lines within each plot indicates substantial uncertainty within individuals and between individuals (note the differences between the three panels). The curves in red are the empirical quantile functions computed from the original days of physical activity data. Notice that the x-axis does not begin until about 0.4 because the activity counts are all zero prior to this value. Within-person Wasserstein distances can now be calculated within each individual using each of the integration intervals considered: $I=[0,1]$, $[0.01,0.99]$ and $[0.05,0.95]$. For illustration purposes, we use $J=999$ terms in the Riemann sum approximation. These mutual within-person distances are shown in Figure~\ref{fig:meas}-(C) for each study participant, respecting the convention of showing the same individual on the vertical, stratified by the interval type. 

The within-participant distances for individuals 33274 (first column of Figure~\ref{fig:meas}-(C)) and 40462 (last column of Figure~\ref{fig:meas}-(C)) exhibit similar medians and variability across all interval types. However, the quantile curves for these participants shown in Figure~\ref{fig:meas}-(B) are quite different. In particular, the quantile curves for study participant 40462 have more variability in the low quantile ranges corresponding to probabilities $0.4$ to $0.7$. This indicates that the Wasserstein distances are less sensitive to the low quantile ranges and do not capture this observed type of difference. In contrast, the distance distributions for study participant 34873 have a higher median and variability for intervals $[0,1]$ and $[0.01,0.99]$, though the distribution for interval $[0.05,0.0.95]$ is more comparable with that of the other two study participants. Once again, this emphasizes that the selection of the integration interval can significantly impact distance estimation, even when considering the same study participant (note the decreasing pattern from red to green to blue in the middle panel in Figure~\ref{fig:meas}-(C)). 

These differences in the within-person distance distributions of participant 34873 raises the question of what exactly is different in their quantile functions. It is actually a small detail hidden at the extreme quantiles, corresponding to probabilities larger than $0.9$. Note the tightness of the distribution at the high quantiles for participants 33274 and 40462 in which their quantiles end in a ``pointed'' tail, and contrast that with the same area for study participant 34873 where the quantiles end in a ``fluffier'' tail. Furthermore, note the resulting similarity of the within-person distances for the three interval types for participant 33274 and 40462, yet the decreasing pattern for participant 34873. As we have shown, the Wasserstein distances are highly sensitive to the upper tail of the distribution, which explains the much higher distances and within-person variability for this participant. We contend that a simple inspection of Figure~\ref{fig:meas}-(B) would not have immediately indicated these substantial differences in the subsequent within-person distances.

Figure~\ref{fig:meas} Panel C also indicates that the caliper, the definition of what is close, may depend on the interval used for integration. This is consistent with our previous observations that demonstrated substantial variability of distances within individuals and their sensitivity to extreme quantiles. Indeed, the median of mutual distances for subject 34873 and interval $[0,1]$ is larger than $50$, whereas the median for subjects 33274 and 40462 are much closer to $20$. Recall from Table~\ref{maintable} that the point estimators and number of study participants that were matched varied substantially exactly in this range, which is consistent with observed within-person differences. Another interesting observation is that most within-person distances are larger than $10$ for all three study participants. This indicates that a caliper less than $10$ would be too aggressive and would match individuals who are more alike than they are to themselves; this issue may occur due to the day-to-day variability in physical activity. This analysis indicates that a caliper in the $10$ to $20$ range is too small for matching on PA.

\begin{figure}[H]
    \centering
    \includegraphics[width=0.9\textwidth]{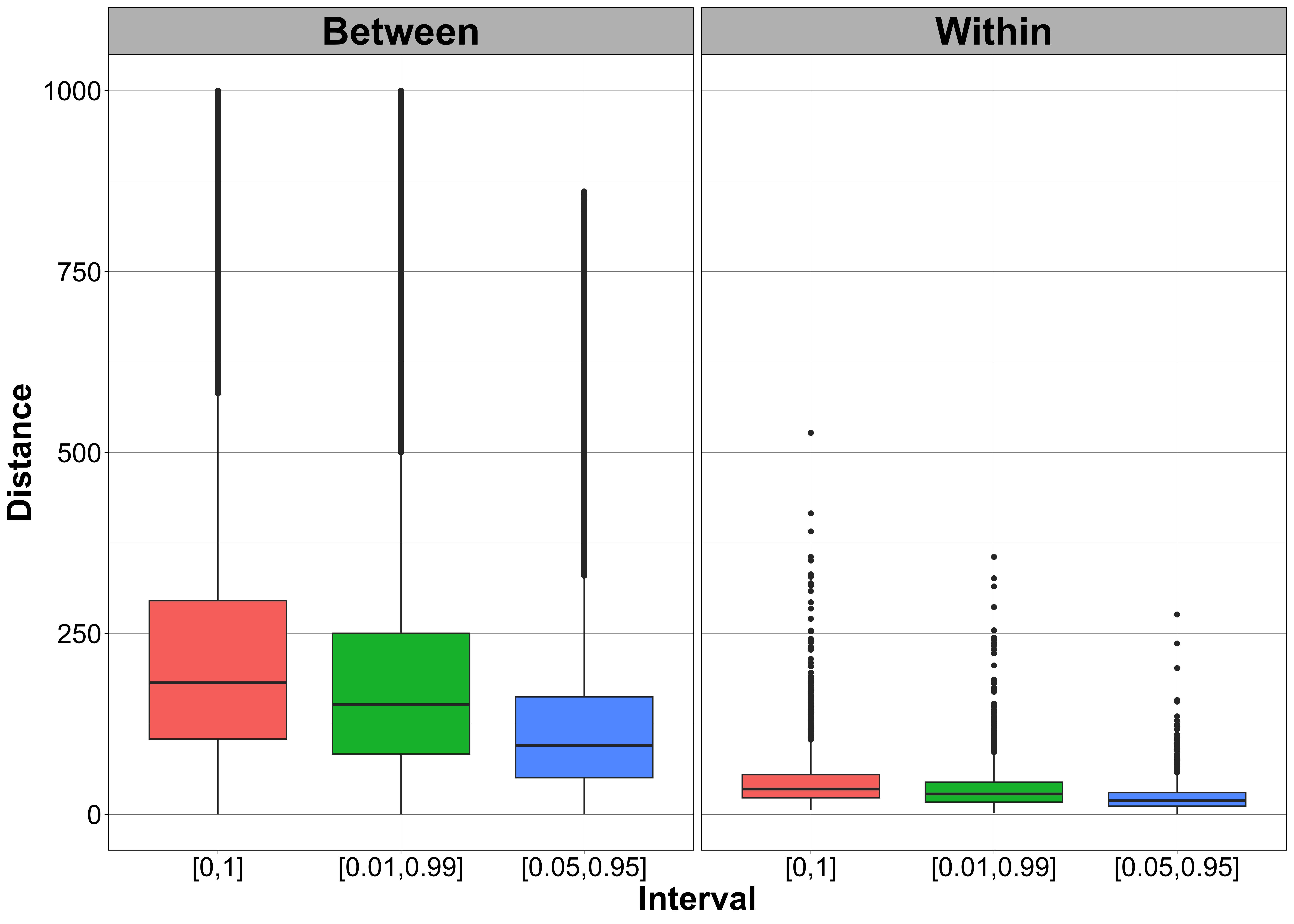}
    \caption{Between Panel: Pairwise distances between PA for all 2486 participants. Within Panel: Within-person distances between 100 resampled quantile functions for each of the 2486 participants.}
    \label{fig:withinbetween}
\end{figure}

Figure~\ref{fig:meas} contains just three participants and it is used to provide the intuition behind the complex decisions that are required in this context. To see whether these findings are reflected in the wider population, we calculate the within- and between-study participants distances for all study participants. Results are displayed in Figure~\ref{fig:withinbetween} separated by between (left panel) and within (right panel) distances. The between-subjects distances are calculated by the Wasserstein distance between the observed empirical quantiles for all $2{,}486$ study participants regardless of race and are displayed as boxplots for each integration interval. The within-subjects distances are calculated by conducting $100$ bootstrap re-samples for each study participant, obtaining the $4{,}950$ distances between these resamples, and calculating the median of these distances. The boxplot of these median within-person distances are displayed in the right panel of Figure~\ref{fig:withinbetween}.

Figure~\ref{fig:withinbetween} indicates that between-participant distances are much larger than within-participant distances, which is expected and reassuring (note that there are distances not shown beyond 1000). There is a declining pattern observed in the median, spread, and skew of the distance distributions across intervals in both the within and between panels, though it is notably more pronounced in the between-panel. This agrees with the intuition that the Wasserstein distance distance is highly sensitive to the high quantiles of the distributions, which in turn are highly affected by measurement error. One could draw the easy conclusion that using the interval $[0.05,0.95]$ addresses some of these problems. A positive aspect could be that, by using this interval, distances decrease; however, there would also be a noticeable rise in the overlap between the within- and between-subject distributions. 

In summary, by utilizing multiple days of data for individuals, we examined the biological measurement error induced by day-to-day variability on the within-person Wasserstein distances. The analyses in Figure~\ref{fig:meas} highlighted the sensitivity of these distances to the choice of integration interval and extreme quantiles of the distributions. This provided context for what is close or far relative to the natural day-to-day variability of physical activity. Figure~\ref{fig:withinbetween} illustrates the within- versus between-Wasserstein distances and provides strong evidence that these distances are indeed very sensitive to the the extreme quantiles of the distributions. This is a problem because, in general, these quantiles are subject to more extreme measurement error, both due to sampling and biological day-to-day variability.

\section{Discussion}\label{sec:discussion}

This paper emphasizes that extensive sensitivity analyses should be conducted in observational studies that conduct matching using high-dimensional covariates. By utilizing these sensitivity analyses, we found that the empirical Wasserstein distance between distributions is highly sensitive to: (1) extreme quantiles of the distributions, which in turn are strongly affected by small sample variability (fewer observations are available to estimate extremes) and biological variability (extreme quantiles have much more day-to-day variability than medians); (2) the practical interval used for integration in the Wasserstein distance, which completely changes the metric of the high-dimensional space (see, for example, Figure~\ref{fig:withinbetween}); and (3) the biological day-to-day variability cannot be ignored and may, in fact, inform the choice of the caliper for deciding what individuals are close or far in terms of high-dimensional distribution. Ignoring these factors can lead to substantial differences between the results of analyses conducted by different research groups on the same data.

Although these problems were identified in the setting of matching on high-dimensional data, extensive sensitivity analyses should be the primary analytic approach for any matching analyses. Indeed, even in low-dimensional settings with perfectly measured covariates, matching on observed covariates does not guarantee balance on unobserved covariates. Moreover, despite the many methods for sensitivity analyses, the current accepted practice does not require sensitivity analyses. This creates a situation where sensitivity analysis is acknowledged as crucial yet seldom implemented. 

This paper illustrates that some of the problems identified for matching in low-dimensional cases are compounded in high-dimensional data analyses. Understanding the data, the implications, and the tuning parameters that make analyses fragile requires extensive knowledge of the subject matter as well as of the technical choices associated with the data analysis.

\newpage

\bibliographystyle{plain}
\bibliography{ref.bib}

\newpage
\section{Supplementary Materials}

\begin{figure}[H]
    \centering
    \includegraphics[width=0.75\textwidth]{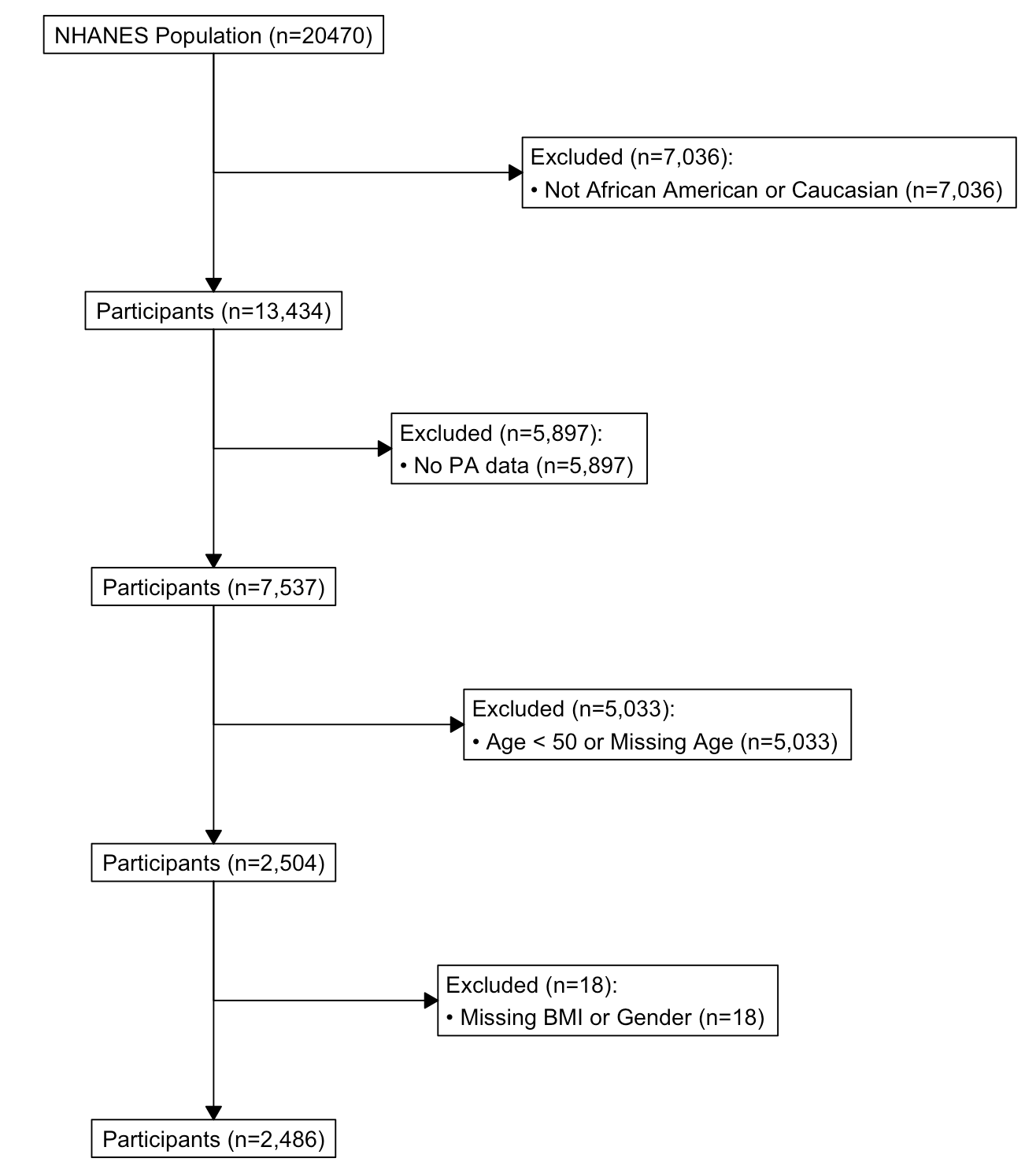}
    \caption{Flowchart illustrating the inclusion/exclusion criteria for participant selection from the 2003-2006 waves of NHANES}
    \label{fig:flowchart}
\end{figure}

\end{document}